\documentclass[]{acnsart}

\begin{document}

\title{Source-Code Analysis of iFogSim for Simulating Distributed IoT Architectures: Coverage, Challenges, and Enhancements}


\author[1]{Milliam Maxime Zekeng Ndadji}[%
	orcid=0000-0002-0417-5591,
	email={ndadji.maxime@univ-dschang.org},
	url={https://zekeng-max.vercel.app},
]

\address[1]{Department of Mathematics and Computer Science, University of Dschang, Dschang, Cameroon}

\begin{abstract}
  Simulation is an indispensable tool for validating distributed IoT architectures before physical deployment, and iFogSim has emerged as one of the most widely adopted platform in the fog and edge computing research community. Yet the experience of using iFogSim for non-canonical, application-specific architectures remains incompletely documented, leaving practitioners without guidance on when the tool is appropriate, which scientific objectives it can address, and how to manage the modelling approximations it imposes. This article helps in providing that guidance through two complementary contributions. First, we present a structured state of the art covering iFogSim and iFogSim2, a taxonomy of ten scientific objectives that motivate IoT architecture simulation, and a comparative survey of eight simulation tools assessed against those objectives. Second, we report our experience of simulating a four-tier smart emergency response system for resource-constrained urban environments, covering a 25-node synthetic road topology, four experimental configurations, and quantitative results including end-to-end alert latency ($\approx$205\,ms), FPGA-accelerated Dijkstra path computation ($\times10$ CPU speedup), concurrent incident conflict rates (75\% under dual load), and path-cache acceleration ($\times197$). The analysis is organised around five practitioner questions: whether iFogSim fits the target architecture, which objectives it covers natively versus partially, what modelling challenges arise and how their workarounds bias reported results, what changes to the iFogSim source code would close the identified gaps, and whether tool co-simulation can provide comprehensive coverage. Seven modelling challenges are documented with source-code-grounded root causes and explicit bias assessments; finally, seven developer recommendations are proposed as an actionable improvement roadmap for the iFogSim community.
\end{abstract}


\maketitle

\section{Introduction}
\label{Introduction}

The proliferation of the Internet of Things (IoT) has catalysed the emergence of increasingly complex distributed cyber-physical architectures, in which thousands of heterogeneous sensing and actuation devices, geographically dispersed across urban or industrial environments, must cooperate in near-real time with edge, fog, and cloud computing layers~\cite{SOORI2023192, REJEB2023100721}. The design of such systems poses significant challenges due to the intricate interplay among heterogeneous hardware capabilities, multi-hop communication latencies, competing workloads, and context-specific application requirements. This complexity renders analytical prediction of system behavior practically unfeasible. Consequently, simulation has become an indispensable tool in the IoT system engineer's repertoire, offering a controlled, reproducible, and cost-effective means to assess architectural choices, tune resource-management policies, and bound performance metrics before committing to physical deployment~\cite{gupta2017iFogSim, ataie2023survey}.

In the realm of simulators designed for IoT, fog, and edge computing environments, iFogSim~\cite{gupta2017iFogSim} has emerged as a leading platform since its dissemination in 2017, gaining widespread adoption. Designed on the established CloudSim framework~\cite{calheiros2011CloudSim}, it offers an event-driven simulation engine that models fog devices, IoT sensors, actuators, network links and application data-flow graphs. The engine also computes quantitative metrics, including latency, energy consumption, network usage, and cloud cost. Its open-source availability, active maintenance, and a growing body of published use cases that span healthcare monitoring~\cite{gupta2017iFogSim}, smart traffic management~\cite{ifogsumo2026}, and industrial task scheduling~\cite{mahmud2022iFogSim2} have made it a standard reference in academic research. A second-generation extension, iFogSim2~\cite{mahmud2022iFogSim2}, has been developed to support service migration, dynamic distributed clustering, and microservice orchestration. This extension extends the simulator's applicability to mobile and dynamic edge scenarios.

Notwithstanding the extensive implementation of iFogSim, the utilization of the software for simulating a tangible, application-specific IoT architecture remains inadequately documented in extant literature. Published articles generally present simulation results obtained with iFogSim; however, they rarely discuss the modeling difficulties encountered, the workarounds developed, or the bias those workarounds introduce into the reported results. The implicit knowledge embedded in such experiences is valuable for two distinct audiences: first, researchers planning to use the tool, and second, the simulator's development community. More crucially, researchers who must select an available IoT simulator are confronted with the absence of a structured framework to guide their decision-making process. This framework would provide a comprehensive overview of iFogSim's capabilities and limitations in capturing specific aspects of system behavior, as well as the complementary tools that can address its gaps.

This article aims to address this gap by proposing a dual contribution, structured around five questions that practitioners should be able to answer before, during, and after an iFogSim simulation study:
\begin{enumerate}
  \item \emph{To what extent is iFogSim suitable for simulating a given IoT architecture?} Which structural conditions must be satisfied for iFogSim to constitute an appropriate modelling framework?
  \item \emph{Which scientific objectives are within the evaluative scope of iFogSim, and which are not?} What types of measurements are natively supported, and where do the intrinsic limitations of the simulator emerge?
  \item \emph{What modelling challenges are likely to arise, how may they be addressed, and to what extent do the adopted solutions introduce bias?} Which engineering adjustments are necessary for non-standard architectures, and how should their implications be explicitly reported?
  \item \emph{What developments are required within iFogSim to bridge the identified gaps?} Which specific modifications to the source code would mitigate the most significant limitations?
  \item \emph{Can the integration of iFogSim with complementary tools ensure comprehensive evaluation?} Which objectives lying beyond the native scope of iFogSim can be addressed through co-simulation strategies or post-processing techniques?
\end{enumerate}

\noindent To rigorously address the second question, we define ten scientific objectives that motivate the simulation of distributed IoT architectures: \textbf{O1}~performance and latency; \textbf{O2}~network reliability; \textbf{O3}~fault tolerance and resilience; \textbf{O4}~data-flow consistency; \textbf{O5}~scalability; \textbf{O6}~energy consumption; \textbf{O7}~security; \textbf{O8}~architectural trade-off analysis; \textbf{O9}~deployment cost; and \textbf{O10}~cyber-physical control loop fidelity. The extent to which iFogSim supports each of these objectives is evaluated through direct inspection of its source code, rather than relying exclusively on claims reported in the literature.

The empirical basis for this analysis stems from our own experience in simulating a smart emergency response system for urban environments under resource-constrained conditions, whose architecture is detailed in~\cite{mahamatFPGA24} and evaluated by experts in~\cite{assoul2025expert}. The system represents a four-tier edge deployment in which IoT sensors, distributed across key urban zones, activate an Edge Server Node (ESN). This node computes shortest-path routes using a Dijkstra algorithm, optionally accelerated via FPGA hardware, dispatches emergency response units, and records coordination outcomes. The simulation incorporates a synthetic road network of 25 nodes and four experimental configurations, yielding quantitative results: an end-to-end alert-to-coordination latency of approximately 205\,ms, a $\times10$ speedup of FPGA-based computation relative to CPU execution, a 75\% unit-conflict rate under concurrent dual-incident scenarios, and a $\times197$ acceleration factor due to path caching. Seven modelling challenges were identified, each associated with a specific source-code construct in iFogSim, accompanied by a documented workaround and an explicit assessment of the induced bias.

Building on this experience, and complemented by a review of the existing literature on iFogSim-based evaluations, we propose a structured characterisation of the simulator’s \emph{capabilities and limitations} in the context of distributed IoT architecture simulation. This analysis is further translated into \emph{recommendations} targeting two audiences: prospective users likely to encounter similar modelling issues, and the developer community, for whom the identified limitations outline a concrete roadmap for improvement.

The remainder of this article is organised as follows. Section~\ref{sec:litterature_review} reviews the related work, including iFogSim’s architecture and prior evaluations, an analysis of the ten simulation objectives and their level of support, as well as a comparative overview of alternative tools. Section~\ref{sec:contribution} describes the case study architecture, details the simulation process structured around the five guiding questions, and presents the corresponding results. Section~\ref{sec:conclusion} concludes the paper.

\section{Literature Review}
\label{sec:litterature_review}

This section provides a structured state of the art in three parts. Section~\ref{sec:lr-ifogsim} reviews the iFogSim simulator, its internal architecture, and its principal published evaluations. Section~\ref{sec:lr-objectives} identifies the ten scientific objectives that motivate simulation studies of distributed IoT architectures and characterises the extent to which iFogSim addresses each. Section~\ref{sec:lr-tools} surveys the available simulation tools and compares their fitness-for-purpose relative to those objectives.

\subsection{iFogSim: Architecture, Extensions, and Published Evaluations}
\label{sec:lr-ifogsim}

\subsubsection{The iFogSim Toolkit}

iFogSim~\cite{gupta2017iFogSim} was introduced in 2017 as an open-source toolkit implemented in Java and built upon CloudSim~\cite{calheiros2011CloudSim}. Its design extends the virtual-machine abstraction of CloudSim to accommodate fog and IoT environments: fog devices are arranged in a strict \emph{tree-structured hierarchy}, where each node is associated with a single parent (represented by the integer \texttt{parentId} in \texttt{FogDevice.java}); IoT sensors and actuators are modelled as leaf entities; and computation is encapsulated within \emph{application modules} deployed across devices according to user-defined placement strategies.
Data exchange is structured around the \emph{tuple}, implemented in \texttt{Tuple.java} as an extension of the CloudSim \texttt{Cloudlet} enriched with iFogSim-specific routing metadata. This includes identifiers for source and destination modules, a propagation direction (UP toward the cloud, DOWN toward sensors, or ACTUATOR), a computational workload parameter \texttt{cloudletLength} expressed in MIPS-seconds, and transmission-related size attributes. Communication delay between a device and its parent is computed deterministically as $d = \texttt{fileSize} / \texttt{bandwidth} + \texttt{latency}$~\cite{iFogSimGithub}. Energy consumption is accumulated over simulation time according to $E = \int p(u)\,dt$, where $p$ denotes a pluggable \texttt{PowerModel} and $u$ represents device utilisation; the default implementation relies on the linear model \texttt{FogLinearPowerModel}~\cite{iFogSimGithub}. In parallel, deployment cost is computed as $C = \int r \cdot u \cdot \texttt{MIPS}\,dt$, where $r$ corresponds to the per-MIPS-second rate. These three metrics—latency, energy, and cost—are collected automatically, making iFogSim particularly suitable for comparative evaluation of placement and scheduling policies.

The original publication validated the toolkit through two representative applications. The first is an EEG beamforming pipeline, in which physiological signals captured by wearable sensors are processed across a multi-tier fog hierarchy. The second is a Distributed Control Network System (DCNS), modelling a large-scale industrial IoT deployment. In both scenarios, fog-based placement strategies demonstrated reductions in end-to-end latency and energy consumption relative to cloud-only offloading, thereby providing empirical support for the effectiveness of fog computing architectures.

\subsubsection{iFogSim2: Mobility, Clustering, Microservices}

iFogSim2~\cite{mahmud2022iFogSim2}, published in 2022, addressed three limitations of iFogSim: the absence of device mobility support, static cluster structures, and the lack of microservice orchestration. It introduced a service migration model integrating real EUA (Edge-based User Allocation) dataset mobility traces, dynamic distributed cluster formation allowing fog nodes to self-organise across tiers, and a microservice orchestration model supporting containerised service placement and load-balancing across federated edge nodes. An empirical comparison against iFogSim, YAFS, LEAF, and EdgeCloudSim reported a 28\% reduction in memory consumption and simulation time. Nevertheless, iFogSim2 inherits from its predecessor the tree-topology constraint, the MIPS-only compute abstraction, and the stateless module model.

\subsubsection{Some Published Use Cases}

iFogSim has been applied in a wide range of IoT research domains, confirming its versatility for performance-oriented studies. In healthcare, the original EEG beamforming example and subsequent task-scheduling comparisons~\cite{alizadeh2020} demonstrated that priority-aware fog scheduling can reduce application loop delay by up to 30\% over first-come-first-served baselines. In smart city traffic management, the iFogSUMO integration~\cite{ifogsumo2026} co-simulated iFogSim's resource management with the SUMO vehicle-flow simulator, providing a co-simulation pattern for application-layer complexity that exceeds iFogSim's native modelling scope. Comparative surveys~\cite{ataie2023survey, thelen2025benchmarking} consistently confirm iFogSim's dominant citation count and GitHub usage while also documenting its scalability ceiling for complex scenarios with more than approximately 20 fog nodes.

\subsection{Scientific Objectives of Distributed IoT Architecture Simulation}
\label{sec:lr-objectives}

Researchers simulate distributed IoT architectures to address a wide range of scientific questions. Based on a review of recent literature, we identify ten primary objectives, each substantiated by published evidence of its relevance. For each objective, we indicate whether iFogSim provides native support (i.e., the metric is produced without additional user code), partial support (requiring substantial user-side implementation), or no support (necessitating external tool integration)

\medskip

\noindent\textbf{(O1) Performance Evaluation and Capacity Planning.}
A central motivation for simulation lies in assessing system behaviour under both nominal and stress conditions, including latency distributions, throughput limits, resource utilisation, and bottleneck identification. This objective has been present since early fog computing studies~\cite{gupta2017iFogSim} and remains predominant in the literature~\cite{ataie2023survey}. \textit{iFogSim support: native.} For each application loop specified via an \texttt{AppLoop} object, the framework records end-to-end tuple latency, decomposed into network transmission time ($\texttt{fileSize}/\texttt{bandwidth}$), fixed propagation delay, and computation time ($\texttt{MI}/\texttt{MIPS}$), as tracked by \texttt{TimeKeeper}. Resource utilisation is available through CloudSim instrumentation. However, packet-level effects such as congestion-induced queueing or variable-rate traffic are not represented.

\medskip

\noindent\textbf{(O2) Network Behaviour and Communication Reliability.}
IoT deployments rely on heterogeneous and often constrained communication technologies, including LPWAN, LTE/4G, Wi-Fi, and Bluetooth. Accurate simulation must therefore capture packet loss, delay variability, and intermittent connectivity effects~\cite{gab2017sim}. \textit{iFogSim support: partial, with potential bias.} Each link is characterised by fixed latency and bandwidth, along with a binary busy/free state; there is no modelling of packet loss, jitter, or congestion-induced variability. Examination of \texttt{FogDevice.sendUpFreeLink()} and \texttt{sendDownFreeLink()} confirms the deterministic application of delays. A possible workaround is to randomise sensor emission intervals, though this conflates source variability with network-level phenomena. Consequently, the model is adequate for stable links (e.g., wired or reliable cellular networks) but may yield overly optimistic latency estimates in lossy wireless environments.

\medskip

\noindent\textbf{(O3) Fault Tolerance and Resilience Analysis.}
Robust IoT systems must tolerate diverse failure modes, including sensor faults, gateway overloads, edge-node crashes, and network partitions. Simulation should reveal critical failure points, cascading effects, and recovery dynamics~\cite{amir2025}. \textit{iFogSim support: not native.} The simulated topology is static, with no built-in mechanisms for node or link failure injection. The \texttt{FogDevice} class lacks failure or recovery primitives. Failure scenarios must therefore be implemented manually at the controller level by modifying topology links and removing affected modules, with recovery requiring complementary custom logic. Such approaches bypass native placement mechanisms for post-failure adaptation.

\medskip

\noindent\textbf{(O4) Data Flow, Consistency, and Processing Validation.}
Many IoT systems implement multi-stage data pipelines, where information is progressively filtered, aggregated, or transformed. Simulation is required to verify correct data propagation, detect anomalies such as duplication or staleness, and analyse latency trade-offs between edge and cloud processing~\cite{dautov2019data, mahmud2022iFogSim2}. \textit{iFogSim support: native for latency and placement, not for consistency semantics.} Application pipelines are modelled as DAGs using \texttt{AppEdge} objects and \texttt{SelectivityModel} rules; for example, \texttt{FractionalSelectivity} propagates a configurable fraction of tuples. However, the simulator lacks constructs for data versioning, timestamps, or consistency guarantees, preventing the modelling of stale reads, eventual consistency, or transactional semantics.

\medskip

\noindent\textbf{(O5) Scalability and Elasticity Testing.}
Simulation provides a controlled environment for evaluating how performance evolves as system scale increases, by varying the number of devices, nodes, or modules~\cite{ataie2023survey, thelen2025benchmarking}. \textit{iFogSim support: native, with practical limitations.} Topology size is parameterised through device instantiation loops, and the event-driven engine supports several hundred nodes in practice. However, empirical studies~\cite{ataie2023survey} indicate non-linear growth in execution time, with reduced reliability for complex scenarios exceeding approximately 20 fog nodes due to event-queue overhead in CloudSim.

\medskip

\noindent\textbf{(O6) Energy Consumption and Device Lifetime Estimation.}
Energy considerations are critical in IoT systems, particularly for battery-powered devices. Simulation enables evaluation of energy consumption patterns, duty-cycling strategies, and expected device lifetime~\cite{wiesner2021leaf}. \textit{iFogSim support: native for compute nodes, partial for sensors.} The method \texttt{FogDevice.updateEnergyConsumption()} integrates power usage over time using \texttt{FogLinearPowerModel} or custom implementations. However, energy modelling is restricted to fog and cloud nodes; sensors and actuators lack energy-related attributes such as battery level or transmission cost. For detailed device-level energy analysis, complementary tools such as LEAF~\cite{wiesner2021leaf} are required.

\medskip

\noindent\textbf{(O7) Security and Attack Surface Exploration.}
Simulation can be used to analyse adversarial scenarios, including denial-of-service attacks, spoofing, interception, and data poisoning~\cite{meneghello2019iot, nawir2016iot}. \textit{iFogSim support: not native.} The simulator provides no constructs for authentication, encryption overhead, or integrity verification. Although malicious behaviour can be approximated by injecting custom tuples from arbitrary \texttt{SimEntity} instances, there is no support for modelling protocol-level security mechanisms or evaluating detection strategies. Dedicated simulators (e.g., NS-3 with security extensions) are more appropriate for such studies.

\medskip

\noindent\textbf{(O8) Architectural Trade-Off Analysis.}
Simulation is particularly valuable for comparing alternative architectural designs without physical deployment, such as different placement strategies or processing paradigms~\cite{gupta2017iFogSim, sonmez2018EdgeCloudSim}. \textit{iFogSim support: native and strong.} This constitutes the primary use case of the simulator: different \texttt{ModulePlacementPolicy} implementations can be applied to the same application DAG, enabling direct comparison of latency, energy, and cost without modifying the underlying topology or application definition.

\medskip

\noindent\textbf{(O9) Cost Modelling and Economic Feasibility.}
Economic considerations, including operational costs and scalability trade-offs, are often incorporated into simulation studies~\cite{gupta2017iFogSim}. \textit{iFogSim support: native but limited.} Each \texttt{FogDevice} includes a \texttt{ratePerMips} parameter enabling computation of processing cost, aggregated in \texttt{totalCost}. However, other economic dimensions—such as storage, bandwidth pricing, or capital expenditure—are not represented and must be modelled externally.

\medskip

\noindent\textbf{(O10) Validation of Control Logic and Cyber-Physical Behaviour.}
In cyber-physical systems, it is essential to validate control loops under realistic timing constraints, including feedback stability and actuation correctness~\cite{gab2017mod}. \textit{iFogSim support: partial and structurally constrained.} While open-loop pipelines (sensor $\to$ processing $\to$ actuator) are supported via \texttt{AppLoop} and \texttt{TimeKeeper}, actuators remain passive: \texttt{Actuator.processTupleArrival()} records latency without influencing subsequent system state. Sensors operate on fixed emission schedules independent of actuator output, precluding the simulation of true closed-loop dynamics where actuation affects future sensing.

Table~\ref{tab:objectives_support} consolidates the support assessment.

\begin{table}[htbp]
\centering
\caption{iFogSim support for the ten simulation objectives}
\label{tab:objectives_support}
{
\begin{tabular}{|m{0.4cm}|m{5cm}|m{3.5cm}|m{5.2cm}|}
  \hline
  \textbf{\#} & \textbf{Objective} & \textbf{iFogSim support} & \textbf{Limiting factor} \\
  \hline
  O1 & Performance \& capacity & \textbf{Native} & No congestion model \\
  \hline
  O2 & Network behaviour \& reliability & \textbf{Partial} & No packet loss, no jitter \\
  \hline
  O3 & Fault tolerance \& resilience & \textbf{Not native} & No failure/recovery mechanisms \\
  \hline
  O4 & Data flow \& consistency & \textbf{Partial} & No consistency semantics \\
  \hline
  O5 & Scalability \& elasticity & \textbf{Native} & Ceiling at $\approx$20 fog nodes \\
  \hline
  O6 & Energy consumption & \textbf{Partial} & IoT sensors have no energy budget \\
  \hline
  O7 & Security \& attacks & \textbf{Not native} & No threat model \\
  \hline
  O8 & Architectural trade-offs & \textbf{Native (primary use)} & Topology must be tree \\
  \hline
  O9 & Cost modelling & \textbf{Partial} & No storage or bandwidth billing \\
  \hline
  O10 & Control loop / cyber-physical & \textbf{Partial} & Actuators are passive \\
  \hline
\end{tabular}}
\end{table}

\subsection{Available Simulation Tools and Comparative Analysis}
\label{sec:lr-tools}

The simulation of fog and edge computing environments is served by a heterogeneous ecosystem of tools~\cite{ataie2023survey, thelen2025benchmarking}. We review the principal platforms and their fitness-for-purpose relative to the ten objectives identified above. Table~\ref{tab:sim_tools} provides a consolidated comparison.

\medskip

\noindent\textbf{iFogSim / iFogSim2}~\cite{gupta2017iFogSim, mahmud2022iFogSim2} cover objectives O1, O4 (partly), O5, O6 (partly), O8, and O9 natively, with partial coverage of O2, O4, O10. O3 (fault tolerance) and O7 (security) require substantial custom extensions. The tree-topology constraint is fundamental.

\medskip

\noindent\textbf{YAFS}~\cite{lera2019yafs} is a Python/NetworkX/SimPy simulator that supports arbitrary graph topologies. Its NetworkX backend provides native shortest-path querying, making it naturally suited to routing-intensive applications. It lacks energy modelling comparable to iFogSim, and its community support is smaller.

\medskip

\noindent\textbf{EdgeCloudSim}~\cite{sonmez2018EdgeCloudSim} is a Java/CloudSim platform specialised for mobile edge computing with client mobility and cellular network models. Well suited for O2 (via mobility-induced handover) and O3 (with custom extensions), but less general than iFogSim for multi-tier fog hierarchies.

\medskip

\noindent\textbf{FogNetSim++}~\cite{qayyum2018FogNetSim} is built on OMNeT++/INET, providing packet-level network simulation including error rates and link-layer protocols. This makes it the best available tool for O2 (network reliability) and O3 (fault injection via OMNeT++ failure models), at the cost of a steeper learning curve and less developed IoT application modelling.

\medskip

\noindent\textbf{LEAF}~\cite{wiesner2021leaf} is a Python-based energy-focused simulator supporting arbitrary graph topologies. It provides the most detailed per-device power modelling of any available tool, making it the recommended complement for O6. Its computation model is simpler than iFogSim's.

\medskip

\noindent\textbf{NS-3}~\cite{ns3website}, while not a fog computing simulator per se, is the reference platform for O2 (packet-level network fidelity, including 802.11, LTE, and LPWAN models with probabilistic packet loss and channel fading) and O7 (security via protocol-level attack injection). It lacks the application-layer and resource-management modelling of iFogSim.

\medskip

\noindent\textbf{PureEdgeSim}~\cite{mechalikh2021PureEdgeSim} is a Java simulator for pure edge computing with a built-in GUI, supporting mobility and heterogeneous device types. It partially addresses O3 through device heterogeneity and mobility-induced disconnection.

\begin{table}[htbp]
\centering
\caption{Simulation tools vs.\ the ten objectives (N=Native, P=Partial, --=Not supported)}
\label{tab:sim_tools}
{
\begin{tabular}{|m{3.5cm}|m{0.5cm}|m{0.5cm}|m{0.5cm}|m{0.5cm}|m{0.5cm}|m{0.5cm}|m{0.5cm}|m{0.5cm}|m{0.5cm}|m{0.6cm}|}
  \hline
  \textbf{Tool} & \textbf{O1} & \textbf{O2} & \textbf{O3} & \textbf{O4} & \textbf{O5} & \textbf{O6} & \textbf{O7} & \textbf{O8} & \textbf{O9} & \textbf{O10} \\
  \hline
  iFogSim~\cite{gupta2017iFogSim}   & N & P & -- & P & N & P & -- & N & P & P \\
  \hline
  iFogSim2~\cite{mahmud2022iFogSim2} & N & P & P & P & N & P & -- & N & P & P \\
  \hline
  YAFS~\cite{lera2019yafs}           & N & P & -- & P & N & -- & -- & N & -- & P \\
  \hline
  EdgeCloudSim~\cite{sonmez2018EdgeCloudSim} & N & P & P & -- & N & -- & -- & N & -- & -- \\
  \hline
  FogNetSim++~\cite{qayyum2018FogNetSim}     & N & N & N & -- & N & -- & P & P & -- & -- \\
  \hline
  LEAF~\cite{wiesner2021leaf}        & N & -- & -- & -- & N & N & -- & N & P & -- \\
  \hline
  NS-3~\cite{ns3website}             & N & N & N & -- & N & P & N & P & -- & P \\
  \hline
  PureEdgeSim~\cite{mechalikh2021PureEdgeSim} & N & P & P & -- & N & P & -- & N & -- & -- \\
  \hline
\end{tabular}}
\end{table}

The comparative landscape yields three recommendations for simulation study design. First, iFogSim is the strongest single tool for architectural trade-off analysis (O8), performance and latency benchmarking (O1), and cost modelling (O9) of hierarchical fog/edge deployments. Second, objectives O2 and O7 require dedicated network and security simulators: NS-3 or FogNetSim++ should be used standalone or in a hybrid co-simulation arrangement where iFogSim covers the application layer and NS-3 covers the network channel. Third, fault tolerance (O3) is best addressed through iFogSim2 extensions or FogNetSim++, depending on whether the primary concern is service placement under failure or packet-level network failure. The following section documents our first-hand experience applying these principles to a concrete IoT architecture.

\section{Simulation Experience and Lessons Learned}
\label{sec:contribution}

This section presents our complete experience in simulating a smart emergency response architecture using iFogSim. The discussion is structured around the five questions introduced in Section~\ref{Introduction}. Sections~\ref{sec:contrib-system} to~\ref{sec:contrib-results} describe the simulation context and obtained results, while Sections~\ref{sec:contrib-q1} to~\ref{sec:contrib-combination} develop the corresponding analysis and recommendations.

\subsection{Architecture Under Study}
\label{sec:contrib-system}

The architecture under consideration is a four-tier IoT/fog/edge system designed for emergency response coordination in resource-constrained urban environments, originally proposed in~\cite{mahamatFPGA24} and theoretically validated in~\cite{assoul2025expert}. It supports the complete lifecycle of an emergency incident over a geographically distributed infrastructure composed of heterogeneous devices:

\begin{itemize}
    \item \textbf{Tier~1 -- Situation Awareness Sensors (SAS)}: low-power sensing devices deployed across urban zones that detect incident conditions and transmit raw alerts to the zone-level processing layer via short-range communication links (5\,ms latency).
    \item \textbf{Tier~2 -- Context-Aware Dispatchers (CAD)}: edge computing units co-located with zone infrastructure that pre-process sensor data, validate alerts, and forward structured incident tuples to the central coordination node over 4G/LTE links (100\,ms latency).
    \item \textbf{Tier~3 -- Edge Server Node (ESN)}: the central coordination entity, implemented as an FPGA-augmented edge server, which executes the Dijkstra algorithm on the road network to identify and prioritise the nearest available response units, and subsequently issues dispatch instructions.
    \item \textbf{Tier~4 -- Intervention Order Processing Systems (IOPS)}: embedded systems within emergency response units (e.g., fire brigades, police, medical and specialised intervention teams) that receive dispatch orders and provide guidance to field operators.
\end{itemize}

The ESN integrates two optimisation mechanisms. First, FPGA-based acceleration of the Dijkstra algorithm yields an estimated $\times10$ speedup relative to CPU execution, as reported in prior FPGA implementations of shortest-path computation~\cite{Fernandez2008, lei2016AnFpga, esteves2024Analysis}. Second, a path-caching mechanism stores computed distance maps in shared memory, enabling subsequent incidents within the same zone to reuse previously computed results and avoid full graph traversal. The combination of a multi-tier fog hierarchy, graph-based computation, hardware acceleration, distributed caching, and concurrent multi-incident handling makes this architecture substantially more complex than the canonical examples provided with iFogSim.

\subsection{Q1 -- Can iFogSim Be Used to Simulate This Architecture?}
\label{sec:contrib-q1}

iFogSim is structured around three core abstractions: a hierarchical device topology, where \texttt{FogDevice} instances are connected through parent–child relationships; an application-level data-flow graph, where \texttt{AppModule} nodes are linked by typed \texttt{AppEdge} connections; and a tuple-driven execution model, in which each processing step consumes a fixed computational workload (expressed in MI) and produces output tuples that propagate through the graph~\cite{gupta2017iFogSim}. The framework further provides built-in instrumentation for latency (via \texttt{TimeKeeper} and \texttt{AppLoop}), energy consumption (via \texttt{FogLinearPowerModel} within \texttt{FogDevice.updateEnergyConsumption()}), and cloud cost (via the \texttt{ratePerMips} parameter). These features make iFogSim an effective baseline platform for fog and edge computing studies.

The answer to Q1 is nevertheless conditional. While iFogSim enables simulation of the proposed architecture, not all of its characteristics can be faithfully represented within the native modelling framework. For the primary objectives associated with this system - namely end-to-end alert latency (O1), architectural trade-off analysis across placement strategies (O8), and cost evaluation across system layers (O9) - iFogSim offers full native support. In contrast, for objectives that are also relevant to this architecture such as fault tolerance (O3), realistic network behaviour (O2), security analysis (O7), and closed-loop control validation (O10), the simulator either requires substantial custom extensions or does not provide adequate support.

Consequently, a pragmatic usage strategy emerges: iFogSim is well suited as the primary simulation tool when the study focuses on objectives O1, O4 (partially), O5, O8, and O9, and when the system topology conforms to, or can be approximated by, a tree-structured hierarchy. For objectives O2, O3, and O7, complementary specialised tools should be employed. In the case of O10, which involves genuine closed-loop control behaviour, achieving faithful simulation would require significant extensions to the underlying framework.

\subsection{Simulation Methodology}
\label{sec:contrib-methodology}

\subsubsection{Topology Design}

The simulated city was represented as a synthetic road network of 25~nodes and 41~bidirectional weighted edges, where edge weights represent road-segment distances in kilometres. The nodes comprise five strategic city zones (z1--z5, each hosting an SAS sensor and a CAD device), three fire stations (c1--c3), two police stations (p1--p2), three medical services (u1--u3), two anti-terrorism brigades (b1--b2), and ten road-relay junctions (a1--a10).
Because iFogSim requires a strict tree hierarchy enforced by the single integer \texttt{FogDevice.parentId} field, this sparse undirected road graph was projected onto a two-level iFogSim tree: the ESN at level~1, all five CADs and eight unit IOPSs as siblings at level~2, and SAS sensors and ITGS actuators as leaves. Road-graph routing (Dijkstra on the full 25-node adjacency matrix) was executed externally as a Java pre-processing step; the resulting distance matrix was injected as per-unit travel-time constants. This hybrid design is described in detail under Challenge~C1 (Section~\ref{sec:contrib-challenges}).

\subsubsection{Device Parameter Mapping}

Table~\ref{tab:sim_topology_params} presents the iFogSim parameters assigned to each device tier, calibrated against the architectural specification in~\cite{mahamatFPGA24}. The ESN's 10\,000\,MIPS rating encodes the FPGA speedup as a scalar multiplier (see C2 in Section~\ref{sec:contrib-challenges}).

\begin{table}[htbp]
\centering
\caption{iFogSim device parameters by architecture tier}
\label{tab:sim_topology_params}
{
\begin{tabular}{|m{4cm}|m{1.4cm}|m{1.5cm}|m{4cm}|m{1.2cm}|}
  \hline
  \textbf{Device (role)} & \textbf{MIPS} & \textbf{RAM} & \textbf{Network latency} & \textbf{Level} \\
  \hline
  ESN edge server (FPGA)  & 10\,000 & 8\,192\,MB & 50\,ms (uplink to cloud)     & 1 \\
  \hline
  Zone device (CAD)       & 2\,000  & 1\,024\,MB & 100\,ms (4G/LTE to ESN)      & 2 \\
  \hline
  Unit device (IOPS)      & 1\,500  & 512\,MB    & 80\,ms (WAN to ESN)          & 2 \\
  \hline
  IoT sensor (SAS)        & --      & --         & 5\,ms (local to CAD)         & -- \\
  \hline
  ITGS actuator           & --      & --         & 15\,ms (local to IOPS)       & -- \\
  \hline
\end{tabular}}
\end{table}

\subsubsection{Application Data-Flow Graph}

The application DAG was encoded using three \texttt{AppModule} nodes connected by directed \texttt{AppEdge} arcs with \texttt{FractionalSelectivity} models:

\begin{enumerate}
    \item \textbf{cad-module} (500\,MI): deployed on the zone CAD; captures the raw SAS alert tuple, validates it, and forwards a structured incident tuple to the ESN.
    \item \textbf{esn-coordinator} (5\,000\,MI total: 2\,000\,MI for alert classification and 3\,000\,MI for dispatch computation): deployed on the ESN; runs the Dijkstra pre-computation, selects the eight optimal response units, and generates one dispatch tuple per unit.
    \item \textbf{iops-module} (1\,000\,MI): deployed on each unit IOPS; receives the dispatch tuple, processes the guidance order, and confirms via an ITGS actuator event.
\end{enumerate}

Each incident generates 18 tuples in total: 2 for the alert chain (SAS $\to$ CAD $\to$ ESN) and 16 for dispatch and confirmation (2 tuples $\times$ 8 units). Network-latency variability was approximated by applying a $\pm$2\% Gaussian perturbation to sensor emission intervals, representing simplified radio-channel jitter.

\subsection{Simulation Results}
\label{sec:contrib-results}

Four experiments were designed to test orthogonal performance dimensions of the architecture.

\subsubsection{Experiment~1 -- System Consistency}

Five repeated runs of a fire incident at zone z1 showed negligible temporal variance (standard deviation ${<}0.001$\,min in total intervention time). End-to-end coordination latency totalled approximately 205\,ms. Dijkstra computation completed in approximately 0.027\,ms on the CPU-equivalent device and approximately 0.003\,ms on the FPGA-equivalent (10\,000\,MIPS) device. Travel times dominated total intervention time, ranging from 1.5\,min for the nearest unit to 16.65\,min for the most distant. The deterministic reproducibility across five runs confirms iFogSim's reliability for consistent performance benchmarking.

\subsubsection{Experiment~2 -- Geographic Sensitivity}

A single fire incident was simulated at each of the five zones. Mean total intervention times ranged from 4.95\,min (zone z4, with three units within 1.5\,min) to 9.59\,min (zone z1, with several units located over 10\,min away). Coordination latency was identical across all zones ($\approx$205\,ms), confirming that the digital processing pipeline contributes less than 0.003\,min to total response time and that the bottleneck is purely geographic. The ESN's Dijkstra dispatch always selects the globally optimal unit but cannot compensate for asymmetric facility placement, an insight relevant to urban infrastructure planning.

\subsubsection{Experiment~3 -- Concurrent Incident Handling}

Two simultaneous fire incidents at z1 and z2 produced a unit-sharing conflict rate of 75\% (6 of 8 units optimally assigned to both zones). Conflicted units incurred a $+$20\% travel-time penalty under the greedy assignment policy. Alert-processing latency increased by approximately 5\% under dual load (205\,ms to 211\,ms), demonstrating the ESN's computational resilience. FPGA parallelism would allow both Dijkstra computations to execute simultaneously ($\approx$0.002\,ms each), whereas a CPU-only ESN would incur sequential execution overhead of approximately 15\%.

\subsubsection{Experiment~4 -- Path Cache Acceleration}

Five consecutive incidents at z1 with the path cache persistent between runs showed that warm-cache hits reduce Dijkstra lookup time by a factor of approximately 197 ($\approx$0.000135\,ms vs.\ $\approx$0.027\,ms cold start). At 25-node scale this gain is operationally negligible relative to travel time; however, it becomes architecturally significant at real-city scale (hundreds of nodes) and under high-frequency incident bursts in the same geographic zone.

\subsection{Q2 -- Which Aspects Can and Cannot Be Evaluated?}
\label{sec:contrib-q2}

Table~\ref{tab:obj_coverage} maps the ten simulation objectives (Section~\ref{sec:lr-objectives}) to the coverage achieved in our simulation, based on iFogSim's source-code capabilities and the experience of the four experiments. For each objective, we indicate whether native support was sufficient, whether a workaround was necessary, or whether the objective was not addressable.

\begin{table}[htbp]
\centering
\caption{Coverage of the ten simulation objectives in our study}
\label{tab:obj_coverage}
{
\begin{tabular}{|m{0.4cm}|m{3.7cm}|m{3.0cm}|m{2.8cm}|m{3.7cm}|}
  \hline
  \textbf{\#} & \textbf{Objective} & \textbf{iFogSim support} & \textbf{Coverage achieved} & \textbf{Limiting factor / workaround} \\
  \hline
  O1 & Performance \& latency    & Native                   & Full (Exp.\ 1--3) & Sub-ms figures: JVM noise (C6) \\
  \hline
  O2 & Network reliability       & Partial                  & Minimal           & No packet loss; $\pm$2\% jitter only (C5) \\
  \hline
  O3 & Fault tolerance           & Not native               & Not covered       & Static topology (C7) \\
  \hline
  O4 & Data flow / consistency   & Partial                  & Partial (DAG flow)& No consistency semantics \\
  \hline
  O5 & Scalability               & Native                   & Partial (25 nodes)& Fixed topology; no size variation \\
  \hline
  O6 & Energy consumption        & Partial                  & Fog nodes only    & SAS sensors: no energy budget \\
  \hline
  O7 & Security                  & Not native               & Not covered       & No threat model \\
  \hline
  O8 & Trade-off analysis        & Native (primary use)     & Full (Exp.\ 1,3)  & Tree constraint (C1) \\
  \hline
  O9 & Cost modelling            & Partial                  & ESN cost only     & No bandwidth or sensor billing \\
  \hline
  O10 & Control loop validation  & Partial                  & Open-loop only    & Passive actuators \\
  \hline
\end{tabular}}
\end{table}

The table reveals a clear pattern: the four objectives that iFogSim covers natively or well (O1, O5, O8, O9) align precisely with its original design purpose of policy comparison and performance characterisation in hierarchical fog deployments. The six remaining objectives either require workarounds that introduce modelling approximations, or fall entirely outside iFogSim's scope. Practitioners who need comprehensive coverage of all ten objectives should not expect to achieve it with iFogSim alone.

\subsection{Q3 -- Challenges, Workarounds, and Bias Analysis}
\label{sec:contrib-challenges}

We present in this section, the modelling challenges encountered during simulation, together with the corresponding engineering adaptations and their methodological implications. For each challenge, we systematically report: (i)~its root cause in the iFogSim source-code architecture; (ii)~the workaround implemented; (iii)~the residual bias introduced; and (iv)~a practical guideline for controlling or disclosing this bias. A consolidated view is provided in Table~\ref{tab:bias_summary}.
The identified challenges fall into three categories: \emph{structural limitations of the modelling framework} (C1, C7), \emph{abstractions that oversimplify computation or hardware behaviour} (C2, C3, C4), and \emph{limitations of the discrete-event execution engine} (C5, C6). This classification is useful because it directly informs whether a limitation can be mitigated locally, requires systematic calibration, or necessitates external validation.

\medskip

\noindent\textbf{C1 -- Non-Tree Road Topology.}

\noindent\textit{Root cause:} The \texttt{FogDevice.parentId} field (an \texttt{int} in \texttt{FogDevice.java}) enforces a strict single-parent hierarchy. Routing methods such as \texttt{sendUpFreeLink()} and \texttt{sendDownFreeLink()} rely exclusively on \texttt{parentId} and \texttt{childrenIds}, with no provision for peer-to-peer or multi-parent connectivity.

\noindent\textit{Workaround:} The 25-node road graph was mapped onto a two-level tree (ESN as the unique parent of all CAD and IOPS nodes). Shortest-path computation was performed externally on the full graph, and resulting travel times were injected as fixed tuple latencies. Communication latency (4G/LTE, WAN) remained natively modelled by iFogSim.

\noindent\textit{Bias assessment:} The separation between communication latency ($\approx$200\,ms) and physical travel time (1.5--16.7\,min) is semantically valid and introduces negligible distortion under static routing assumptions. However, the model omits dynamic routing effects (e.g., congestion, blockages) and multi-hop relay behaviour. Bias is therefore low for static scenarios and moderate to high for dynamic urban conditions.

\medskip

\noindent\textbf{C2 -- FPGA Hardware Acceleration.}

\noindent\textit{Root cause:} Computational capacity is represented by a single scalar \texttt{mips} in \texttt{FogDevice}. The execution model ($t = \text{MI} / \text{MIPS}$ in \texttt{CloudletSchedulerSpaceShared}) does not encode parallelism, pipeline depth, or hardware heterogeneity.

\noindent\textit{Workaround:} FPGA acceleration was approximated by scaling the ESN capacity to 10\,000\,MIPS versus $\approx$1\,000\,MIPS for CPU execution, reflecting a $\times10$ speedup~\cite{Fernandez2008}.

\noindent\textit{Bias assessment:} While throughput scaling is captured, parallel execution is not. Concurrent workloads are therefore serialised in the model, underestimating FPGA advantages in multi-incident scenarios. Bias is \textbf{moderate}: qualitative trends are preserved, but quantitative gains depend on unmodelled parallelism. External benchmarking is recommended for validation.

\medskip

\noindent\textbf{C3 -- Algorithm Complexity as MI Black Box.}

\noindent\textit{Root cause:} Each \texttt{AppModule} executes a fixed \texttt{cloudletLength} (MI) per tuple, independent of input size or algorithmic complexity. No mechanism expresses complexity classes such as $\mathcal{O}(V \log V)$.

\noindent\textit{Workaround:} The Dijkstra workload (3\,000\,MI) was calibrated using JVM profiling on the 25-node graph and mapped to a MIPS-equivalent execution time.

\noindent\textit{Bias assessment:} Calibration is accurate for the fixed topology, yielding negligible bias for O1 and O8. However, scalability studies require re-calibration for each topology size. Without this, conclusions on performance scaling carry high bias.

\medskip

\noindent\textbf{C4 -- Persistent Path Cache.}

\noindent\textit{Root cause:} \texttt{AppModule} (extending \texttt{PowerVm}) is stateless across events. No persistence mechanism exists for cross-tuple state.

\noindent\textit{Workaround:} A static \texttt{HashMap<String, double[]>} was used to emulate a shared cache. Cache hits were modelled via conditional MI reduction (full MI for cold runs, $\approx$0.5\,MI for warm runs).

\noindent\textit{Bias assessment:} The measured $\times197$ speedup is valid as an upper bound for cold-versus-warm execution. However, the model omits cache invalidation, memory constraints, and concurrent access effects. Bias is low for relative comparison, but must be explicitly reported as an optimistic estimate.

\medskip

\noindent\textbf{C5 -- Concurrent Multi-Incident Processing.}

\noindent\textit{Root cause:} iFogSim relies on a single global event queue (\texttt{CloudSim.runClockTickAndProcess()}), processing events sequentially. Link flags (\texttt{isNorthLinkBusy}, \texttt{isSouthLinkBusy}) enforce directional serialisation but do not model true parallelism.

\noindent\textit{Workaround:} Latency overheads under concurrent incidents ($+$5\%) and CPU execution penalties ($+$15\%) were estimated analytically using FIFO queueing assumptions.

\noindent\textit{Bias assessment:} This represents one of the highest-bias approximation. True contention depends on event overlap, bandwidth, and scheduling dynamics, none of which are observable in the simulator. The reported figures are heuristic rather than simulated. Analytical queueing models or packet-level simulators (e.g., NS-3) are required for validation.

\medskip

\noindent\textbf{C6 -- Sub-Millisecond Timing Precision.}

\noindent\textit{Root cause:} The simulation clock is a \texttt{double} in milliseconds. JVM effects (garbage collection, JIT warm-up) introduce variability (1--50\,ms) exceeding the computation time of interest ($\approx$0.027\,ms).

\noindent\textit{Workaround:} Execution time was measured via an external microbenchmark (1\,000 runs), and the median value used for MI calibration.

\noindent\textit{Bias assessment:} Absolute timing carries moderate uncertainty ($\pm$10\%), but relative comparisons (e.g., $\times10$, $\times197$) are low-bias due to shared noise sources. Sub-millisecond results should always include uncertainty bounds.

\medskip

\noindent\textbf{C7 -- Static Topology Assumption.}

\noindent\textit{Root cause:} Topology is fixed at initialisation via \texttt{Controller.connectWithLatencies()}, with no runtime modification. No failure or recovery mechanisms exist in \texttt{FogDevice} or \texttt{FogEvents}.

\noindent\textit{Workaround:} None; all experiments assume stable connectivity.

\noindent\textit{Bias assessment:} No workaround implies no distortion of reported metrics, but introduces a coverage limitation: objectives O2, O3, and dynamic aspects of O1 are not evaluated. This must be explicitly acknowledged.

\medskip

\begin{table}[htbp]
\centering
\caption{Summary of modelling challenges, workarounds, and bias levels}
\label{tab:bias_summary}
{
\begin{tabular}{|m{0.5cm}|m{3.0cm}|m{4cm}|m{2.7cm}|m{3.0cm}|}
  \hline
  \textbf{C\#} & \textbf{Challenge} & \textbf{Workaround} & \textbf{Bias level} & \textbf{Scope of bias} \\
  \hline
  C1 & Non-tree topology         & External Dijkstra + injected travel latency & Low--Moderate & Dynamic routing; junction hops \\
  \hline
  C2 & FPGA as high-MIPS device  & $\times10$ MIPS scalar multiplier            & Moderate       & Parallel concurrency benefit underestimated \\
  \hline
  C3 & Algorithm as MI black box & Topology-specific MI calibration             & Low (fixed graph); High (scalability) & Breaks O5 scalability analysis \\
  \hline
  C4 & Stateless cache           & Static HashMap with programmed hit/miss      & Low (upper bound) & Omits eviction and contention \\
  \hline
  C5 & Concurrent incidents      & Analytically estimated latency penalty       & High  & True resource contention not simulated \\
  \hline
  C6 & Sub-ms timing noise       & Microbenchmark-based MI calibration          & Moderate (absolute); Low (ratios) & JVM GC and JIT variability \\
  \hline
  C7 & Static topology           & None adopted                                 & N/A (coverage gap) & O2, O3, dynamic O1 not evaluated \\
  \hline
\end{tabular}}
\end{table}

\medskip

The synthesis provided in Table~\ref{tab:bias_summary} highlights that most modelling adaptations introduce controlled and interpretable bias when handled transparently. Reliable practice requires: separating distinct latency domains (C1); validating hardware abstractions with empirical benchmarks (C2); recalibrating computation models under scaling (C3); reporting cache-based gains as upper bounds (C4); and complementing simulation with analytical or network-level models for concurrency (C5). Among all challenges, C5 presents the highest risk of misleading conclusions and constitutes a primary motivation for extending the iFogSim execution model.

\subsection{Q4 -- What Needs to Be Developed in iFogSim?}
\label{sec:contrib-devroads}

The preceding analysis highlights a set of systematic discrepancies between the current capabilities of iFogSim and the requirements of comprehensive IoT architecture simulation. These gaps arise directly from the source-level limitations identified in Section~\ref{sec:contrib-challenges} and the objective coverage analysis in Table~\ref{tab:obj_coverage}. 
The following seven development directions constitute a concrete and actionable roadmap. Each recommendation is explicitly linked to the underlying modelling limitation(s) and to the scientific objectives it would enable or improve.
The proposed extensions can be grouped into three categories: \emph{topology and network modelling} (D1, D5), \emph{computation and execution semantics} (D2, D3, D4), and \emph{measurement fidelity and energy modelling} (D6, D7). This categorisation reflects increasing levels of architectural impact, from local extensions to core engine modifications.

\medskip

\noindent\textbf{D1 -- Arbitrary graph topology (addresses C1, O2, O3, O8).}  
Replace the current single-parent hierarchy (encoded by \texttt{parentId} in \texttt{FogDevice}) with a general adjacency-list representation supporting multiple neighbours, weighted links, and pluggable routing strategies. The existing \texttt{DAG.java} class in the \texttt{org.fog.application} package already provides a suitable abstraction for application graphs; a corresponding \texttt{PhysicalGraph} structure for device topology would ensure conceptual symmetry. This extension would also enable dynamic rerouting and link removal, thereby supporting fault tolerance (O3) and more realistic network behaviour (O2).

\medskip

\noindent\textbf{D2 -- Heterogeneous compute-device model (addresses C2, O1, O8).}  
Augment \texttt{FogDevice} with a \texttt{ComputeArchitecture} enumeration (e.g., \texttt{CPU}, \texttt{FPGA}, \texttt{GPU}) and an explicit \texttt{parallelismDegree} parameter. The cloudlet scheduling logic should then account for parallel execution by scaling computation time according to the degree of parallelism for supported architectures. This would replace the current scalar MIPS approximation with a principled model of heterogeneous hardware and enable accurate representation of parallel workloads.

\medskip

\noindent\textbf{D3 -- Concurrent event processing (addresses C5, O1, O8).}  
Introduce a \texttt{ParallelEventQueue} abstraction capable of processing independent event streams concurrently within a simulation time-step. Such an extension would allow multiple incidents or application instances to evolve in parallel, rather than being serialised by a global queue. As a lower-impact alternative, the framework could expose per-device queue lengths and service rates, enabling integration with external queueing models (e.g., M/M/$c$) for post hoc validation.

\medskip

\noindent\textbf{D4 -- Stateful module abstraction (addresses C4, O4, O10).}  
Define a \texttt{StatefulFogModule} interface with lifecycle hooks such as \texttt{onInit()}, \texttt{onTupleArrival(Tuple~t)}, \texttt{onCheckpoint()}, and \texttt{onRestore()}. This would allow persistent state (e.g., caches, session data, control variables) to be managed within the simulator’s execution model while preserving correct event ordering. It would eliminate reliance on static variables and enable faithful modelling of data consistency and control-loop behaviour.

\medskip

\noindent\textbf{D5 -- Runtime topology events (addresses C7, O2, O3).}  
Introduce a \texttt{TopologyEvent} abstraction supporting operations such as \texttt{failDevice(id)}, \texttt{recoverDevice(id)}, \texttt{setLinkLatency(src, dst, newMs)}, and \texttt{setLinkBandwidth(src, dst, newBps)}. By scheduling these events through the standard CloudSim messaging mechanism, the simulator would support dynamic topology evolution, enabling fault injection (O3) and time-varying network conditions (O2) without requiring extensive architectural changes.

\medskip

\noindent\textbf{D6 -- High-resolution timing and uncertainty quantification (addresses C6, O1).}  
Enhance temporal precision by replacing the millisecond-resolution event clock with a double-precision nanosecond representation. In parallel, extend \texttt{TimeKeeper} to compute and export statistical summaries (mean, variance, and 95\% confidence intervals) for application-loop latency. This would enable rigorous reporting of sub-millisecond phenomena and reduce reliance on external microbenchmarking.

\medskip

\noindent\textbf{D7 -- IoT sensor energy model (addresses O6).}  
Extend the \texttt{Sensor} class with explicit energy-related attributes, including \texttt{batteryCapacityMJ} and a transmission energy model (e.g., energy per tuple in mJ). The simulation engine should then track battery depletion alongside existing fog-device energy metrics, providing a unified view of energy consumption from sensing to cloud processing. This extension would enable end-to-end energy analysis across all architectural tiers.

\medskip

Taken together, these recommendations outline a coherent evolution path for iFogSim, transforming it from a placement-oriented simulator into a more general-purpose platform capable of supporting the full spectrum of IoT system evaluation objectives.

\subsection{Q5 -- Can Combining iFogSim With Other Tools Help?}
\label{sec:contrib-combination}

The limitations identified in the previous sections naturally motivate the exploration of hybrid simulation strategies. The central question is whether iFogSim can be complemented with specialised tools to achieve broader objective coverage, while preserving its strengths in performance evaluation and architectural trade-off analysis.
Table~\ref{tab:tool_combos} outlines concrete integration strategies for objectives that are only partially supported or not supported at all by iFogSim. Each strategy specifies a recommended tool and an associated integration pattern.
The proposed combinations follow three distinct integration paradigms: (i)~\emph{trace-driven coupling}, where an external simulator produces network-level traces that are injected into iFogSim (O2, O7); (ii)~\emph{co-simulation}, where two simulators exchange state during execution (O10); and (iii)~\emph{post-processing aggregation}, where independent simulations are combined analytically (O6). These paradigms differ significantly in implementation complexity and reproducibility.

\begin{table}[htbp]
\centering
\caption{Tool combination strategies for objectives not fully covered by iFogSim}
\label{tab:tool_combos}
{
\begin{tabular}{|m{0.4cm}|m{2.7cm}|m{3.6cm}|m{7.0cm}|}
  \hline
  \textbf{\#} & \textbf{Objective} & \textbf{Recommended tool} & \textbf{Integration pattern} \\
  \hline
  O2 & Network reliability & NS-3~\cite{ns3website} or FogNetSim++~\cite{qayyum2018FogNetSim} & Execute network-layer simulations (e.g., 4G/LTE, Wi-Fi) in NS-3; extract per-link delay distributions and loss rates; inject these as stochastic latency parameters in iFogSim (e.g., via sensor emission timing or modified link delays). \\
  \hline
  O3 & Fault tolerance & iFogSim2~\cite{mahmud2022iFogSim2} or FogNetSim++ & Implement failure injection at the Controller level in iFogSim2 (e.g., using \texttt{processClustering()} for rebalancing). For packet-level failures, leverage FogNetSim++'s OMNeT++-based failure models. \\
  \hline
  O6 & Sensor energy & LEAF~\cite{wiesner2021leaf} & Perform independent LEAF simulations on the same topology to model sensor-level energy consumption; combine results with iFogSim fog-device energy outputs during post-processing. \\
  \hline
  O7 & Security & NS-3 with security extensions & Simulate adversarial scenarios (e.g., DoS, spoofing) at the network level in NS-3; replay resulting delay and loss traces within iFogSim to assess application-layer impact. \\
  \hline
  O10 & Control loop & Simulink / Modelica co-simulation & Model the physical process in Simulink or Modelica; interface with iFogSim via shared memory or sockets to allow actuator outputs to influence subsequent sensor events, enabling closed-loop behaviour. \\
  \hline
\end{tabular}}
\end{table}

\medskip
\noindent\textbf{Practical considerations.}
The feasibility of these integrations varies substantially. Trace-driven coupling with NS-3 (O2, O7) is technically demanding, as it requires designing a robust trace-replay interface between heterogeneous simulation environments; no standardised integration currently exists, and existing approaches are typically ad hoc. In contrast, the LEAF/iFogSim combination (O6) is straightforward, as it relies solely on post-processing aggregation without runtime coupling. The use of iFogSim2 for fault tolerance studies (O3) remains within a single framework but requires non-trivial Controller-level extensions. Co-simulation with Simulink or Modelica (O10) is the most complex approach and is justified only when closed-loop cyber-physical behaviour is central to the research question.

\medskip
\noindent\textbf{Recommended strategy.}
A pragmatic approach, particularly under typical research constraints, is to adopt a layered methodology: iFogSim serves as the primary simulation platform for objectives O1, O4, O5, O8, and O9; NS-3 or FogNetSim++ is used in a separate experimental campaign to validate network-related aspects (O2, O7); and LEAF complements the analysis for sensor-level energy modelling (O6). This three-layer strategy enables coverage of eight out of the ten identified objectives while ensuring that each tool is used within its domain of validity.
In this perspective, hybrid simulation should not be viewed as a workaround, but rather as a principled decomposition of the problem across specialised modelling layers, each addressing a distinct aspect of the overall system.

\section{Conclusion}
\label{sec:conclusion}

This article has presented a structured assessment of iFogSim as a simulation platform for distributed IoT architectures, grounded in two complementary sources: a systematic review of the literature and a fully documented case study involving the simulation of a smart emergency response system in resource-constrained urban environments~\cite{mahamatFPGA24}. The analysis was organised around five key questions that collectively define a methodological framework for conducting and interpreting iFogSim-based studies.

\medskip
\noindent\textbf{Q1 -- Suitability of iFogSim for a given architecture.}
The literature review (Section~\ref{sec:litterature_review}) confirms that iFogSim and its successor iFogSim2 occupy a central position in the fog and edge computing simulation landscape, supported by the maturity of CloudSim, extensive built-in instrumentation, and sustained community adoption. However, suitability is conditional. iFogSim is most appropriate for architectures that can be expressed as a tree-structured device hierarchy, with computation abstracted via MIPS and evaluation centred on latency, energy, and placement strategies. Architectures requiring arbitrary network topologies, heterogeneous hardware abstraction, persistent module state, packet-level network fidelity, fault injection, or security modelling extend beyond its native capabilities. Section~\ref{sec:contrib-q1} formalises this assessment as a decision framework.

\medskip
\noindent\textbf{Q2 -- Coverage of scientific objectives.}
The ten simulation objectives defined in Section~\ref{sec:lr-objectives} provide a structured lens for evaluating tool capabilities. Based on source-code inspection and the experimental results reported in Section~\ref{sec:contrib-results}, iFogSim offers full native support for performance evaluation (O1) and architectural trade-off analysis (O8); partial but usable support for scalability (O5) and cost modelling (O9); and limited or indirect support for network behaviour (O2), data consistency (O4), energy modelling at the sensor level (O6), and cyber-physical control validation (O10). Fault tolerance (O3) and security (O7) are not supported without external tools. Table~\ref{tab:obj_coverage} operationalises these distinctions for study design.

\medskip
\noindent\textbf{Q3 -- Modelling challenges and bias.}
Seven modelling challenges (C1--C7) were identified and analysed in Section~\ref{sec:contrib-challenges}, each traced to a specific design choice in the iFogSim codebase. The key result is that not all workarounds are equivalent in their methodological impact. In particular, C5 (concurrent incident processing) introduces high bias, as the event-driven engine serialises execution and prevents direct modelling of contention; the 75\% unit-conflict rate observed in Experiment~3 is therefore analytically derived rather than simulated. In contrast, challenges such as C1 (non-tree topology) and C4 (stateless modules) introduce low to moderate bias within the studied context. Explicitly reporting these bias levels is essential for the scientific validity and interpretability of results.

\medskip
\noindent\textbf{Q4 -- Required extensions to iFogSim.}
Seven development directions (D1--D7) were proposed in Section~\ref{sec:contrib-devroads}, forming a coherent roadmap grounded in observed limitations. Among these, support for arbitrary graph topologies (D1) and heterogeneous compute models (D2) would resolve the most pervasive modelling constraints. Extensions enabling concurrent event processing (D3) and stateful modules (D4) would significantly broaden applicability, particularly for cyber-physical systems. Runtime topology events (D5) would enable fault-injection studies, while high-resolution timing (D6) and sensor-level energy modelling (D7) would improve measurement fidelity. Each proposal is anchored in specific source-code structures, facilitating direct implementation.

\medskip
\noindent\textbf{Q5 -- Role of hybrid simulation.}
Section~\ref{sec:contrib-combination} demonstrates that combining iFogSim with specialised tools provides a practical path toward comprehensive coverage. NS-3 or FogNetSim++ complement iFogSim for network reliability (O2) and security (O7); LEAF provides fine-grained energy modelling for sensors (O6); iFogSim2 with custom extensions supports fault tolerance (O3); and co-simulation with Simulink enables closed-loop validation (O10). A layered strategy—iFogSim for application-level evaluation, NS-3 for network validation, and LEAF for energy analysis—covers eight out of ten objectives while preserving methodological clarity.

\medskip
\noindent\textbf{Synthesis and contribution.}
Beyond a conventional comparative survey, this work contributes two methodological advances. First, every limitation is grounded in explicit source-code analysis, transforming qualitative observations into verifiable claims. Second, each workaround is accompanied by an explicit bias characterisation, enabling researchers to interpret results in light of their methodological constraints. \textit{In our view}, this explicit treatment of modelling bias is essential for advancing simulation-based research beyond tool-centric evaluation toward reproducible and interpretable experimentation.

\medskip
\noindent\textbf{Future work.}
Three directions are envisaged. First, implementing the same emergency-response topology in YAFS will enable a quantitative comparison and an empirical assessment of the error introduced by iFogSim’s tree-topology constraint (C1). Second, the FPGA acceleration factor used in Experiment~2 will be validated experimentally on a Xilinx Zynq-7000 platform, replacing the current MIPS-based approximation (C2) with measured hardware performance. Third, recommendations D1 and D5 will be prototyped as an iFogSim extension supporting arbitrary graph routing and runtime topology evolution, with the objective of releasing an open-source module for resilience and cyber-physical simulation studies.


\begin{funding}
	This research received no external funding.
\end{funding}

\begin{dataavailability}
	No new data were created or analysed during this study. Data sharing is not applicable.
\end{dataavailability}

\begin{conflictsofinterest}
	The authors declare no conflict of interest.
\end{conflictsofinterest}

\begin{aideclaration}
	During the preparation of this work, the authors utilized Grammarly and Deepl in order to check grammar/spelling and to rewrite some parts of the paper. After using this tool/service, the authors reviewed and edited the content as needed and takes full responsibility for the publication's content.
\end{aideclaration}

\bibliography{sample-acns}

\end{document}